\documentclass[11pt]{article}
\usepackage{amsfonts}
\newcommand{\nn}{\nonumber}
\newcommand{\ti}{\ensuremath{\tilde{I}}}
\newcommand{\tj}{\ensuremath{\tilde{J}}}
\newcommand{\tk}{\ensuremath{\tilde{K}}}
\newcommand{\M}{\ensuremath{\mathcal{M}}}

\makeatletter \@addtoreset{equation}{section} \makeatother
\renewcommand{\theequation}{\thesection.\arabic{equation}}

\begin{document}

\begin{titlepage}
\begin{flushright}
CERN-TH/2001-217\\
hep-th/0108094\\
\end{flushright}
\vspace{0.2cm}
\begin{center}
\begin{LARGE}
\textbf{Options for Gauge Groups in Five-Dimensional Supergravity
}\footnote{ Work supported
  in part by the National
Science Foundation under Grant Number PHY-9802510.}
\end{LARGE}\\
\vspace{1.0cm}
\begin{large}
{\bf John Ellis}$^{\dagger}$ \footnote{John.Ellis@cern.ch}, 
{\bf Murat G\"{u}naydin}$^{\dagger\ddagger}$ \footnote{murat@phys.psu.edu}
and
{\bf Marco Zagermann}$^{\ast}$
\end{large}\footnote{zagermann@physik.uni-halle.de} \\
\vspace{.35cm}
$^{\dagger}$ \emph{Theory Division, CERN,
1211 Geneva 23, Switzerland} \\
\vspace{.3cm}
$^{\ddagger}$ \emph{Physics Department,
Pennsylvania State University,
University Park, PA 16802, USA} \\
\vspace{.3cm}
$^{\ast}$ \emph{Fachbereich Physik,
Martin-Luther-Universit\"{a}t Halle-Wittenberg,\\
Friedemann-Bach-Platz 6, D-06099 Halle, Germany}\\

\vspace{.3cm}

\vspace{0.5cm}
{\bf Abstract}
\end{center}
\begin{small}

Motivated by the possibility that physics may be effectively
five-dimensional over some range of distance scales, we study the possible
gaugings of five-dimensional $\mathcal{N}=2$ supergravity.  Using a
constructive approach, we derive the conditions that must be satisfied by
the scalar fields in the vector, tensor and hypermultiplets if a given
global symmetry is to be gaugeable. We classify all those theories that
admit the gauging of a compact group that is either Abelian or
semi-simple, or a direct product of a semi-simple and an Abelian group. In
the absence of tensor multiplets, either the gauge group must be
semi-simple or the Abelian part has to be $U(1)_R$ and/or an Abelian
isometry of the hyperscalar manifold. On the other hand, in the presence
of tensor multiplets the gauge group cannot be semi-simple. As an
illustrative exercise, we show how the Standard Model $SU(3) \times SU(2) 
\times U(1)$ group may be gauged in five-dimensional $\mathcal{N}=2$
supergravity. We also show how previous special results may be recovered
within our general formalism. 

\end{small}

\end{titlepage}

\renewcommand{\theequation}{\arabic{section}.\arabic{equation}}
\section{Introduction}
\setcounter{equation}{0}

There is currently much interest in the possibility that extra dimensions may appear at 
distance scales that are large relative to the inverse of the Planck length $1/M_P \sim 
10^{-33}$~cm or the Grand Unification scale $1/M_{GUT} \sim 10^{-30}$~cm, and possibly 
at scales accessible to experiments. It is therefore important to understand what gauge 
groups and what matter representations are possible in various dimensions and  what 
restrictions on the underlying `Theory of Everything' may be provided by some variant of 
eleven-dimensional M theory.

One particular scenario for extra dimensions is the original proposal that
eleven-dimensional M theory might be compactified on some Calabi-Yau
manifold down to five dimensions~\cite{HW}. The fifth dimension would then
be just a few orders of magnitude larger than the Planck length or the GUT
scale, and five-dimensional supergravity would be the appropriate
effective low-energy field theory over this range of scales. In this
scenario, the $SU(3) \times SU(2) \times U(1)$ gauge fields of the
Standard Model would be restricted to a brane at one end of the fifth
dimension, and there would be another `hidden' gauge group restricted to
another brane at its other end. Subsequently, elaborations with other
gauge groups appearing on intermediate branes have also been
studied~\cite{Ovrut}. 

In all this class of scenarios, a good characterization of the options
available in the effective intermediate five-dimensional
theory~\cite{CY,Gflux} that governs the dynamics in the bulk between the
branes is essential. For example, this effective theory frequently plays
an essential r\^ole in mediating supersymmetry breaking between the brane
on which it originates and the brane where the Standard Model is
localized~\cite{susyX}. 

Analyses of this class of scenarios have been in the context of
five-dimensional supergravity with only {\it Abelian} gaugings~\cite{CY}. 
This assumption was motivated by the fact that the Ho\v{r}ava-Witten
scenario~\cite{HW} yields a gauging of an Abelian isometry of the
universal hypermoduli space, which originates from the non-vanishing $G$
flux in the underlying eleven-dimensional theory~\cite{Gflux,LOSW2}. 
Supplementary motivation came from the more general expectation that the
Standard Model gauge group would be localized on one brane. 

Calabi-Yau manifolds generically do not possess continuous non-Abelian
global symmetries that are candidates for gauging the five-dimensional
supergravity theory. On the other hand, such symmetries may appear at
singular points in the moduli space of Calabi-Yau manifolds, leading to
the possible appearance of enhanced gauge symmetries~\cite{singularities}.
Moreover, non-perturbative M-theory dynamics may favour some alternatives
to Calabi-Yau compactification possessing global symmetries that might be
gauged. 

One should also remain open to the possibility that the $SU(3) \times
SU(2) \times U(1)$ gauge group of the Standard Model might {\it not} be
restricted to a four-dimensional brane in this higher-dimensional space. A
strong argument against the latter possibility seems to be provided by the
excellent agreement of the values of the gauge couplings measured at low
energies with the predictions of supersymmetric gauge theories in four
dimensions~\cite{GR}. However, it has been observed that gauge-coupling
unification is also possible~\cite{DDG}, in some approximation, even if
the Standard Model gauge group extends into a fifth dimension. Therefore
the possibility of such an extension cannot, perhaps, be rejected
absolutely.

For all these reasons, we think it important to characterize what gauge groups may be 
possible in five-dimensional supergravity, and at what price in terms of restrictions on 
the scalar manifold associated (presumably) with the compactification from higher 
dimensions, in particular its global symmetries.

Previous analyses have focussed on five-dimensional supergravity theories
with scalar manifolds in particular symmetry classes. In this paper, we
attempt a systematic classification of all the options for the
five-dimensional gauge group, noting in each case the appropriate
conditions on the corresponding scalar manifolds. As a special case, we
mention how the $SU(3) \times SU(2) \times U(1)$ gauge group of the
Standard Model may be obtained in a suitable five-dimensional supergravity
theory, not that we recommend it for any particular phenomenological
reasons, but simply as an interesting exercise illustrating our general
results.

The outline of this paper is as follows: In Section 2, we recall the
relevant properties of ungauged $\mathcal{N}=2$ supergravity theories in
five dimensions. Our emphasis is on the global symmetry groups, $G$, of
these theories and their `gaugeable' subgroups $K\subset G$. As shown, the
least trivial part of a classification of admissible gauge groups lies in
the classification of the gaugeable isometries of the vector multiplet
moduli space. In Section 3, which constitutes the main part of this paper,
we give such a classification.  To be precise, we classify all those
theories that admit the gauging of a compact group $K$ that is either
Abelian or semi-simple or a direct product of a semi-simple and an Abelian
group. We illustrate our results with the example of $SU(3) \times SU(2)
\times U(1)$ in Section 4, and summarize and draw some conclusions from
our results in Section 5.  Finally, Appendix A contains a few explicit
examples illustrating our general discussion.

\section{Ungauged Five-Dimensional $\mathcal{N}=2$ Supergravity and its
Possible Gaugings}
\setcounter{equation}{0}

Gauged supergravity theories
are  supergravity theories in which some vector fields $A_{\mu}^I$
are coupled to matter fields $\Phi^{A}$ via gauge
covariant derivatives of the form
\begin{equation}\label{covder}
D_{\mu}\Phi^A\equiv \nabla_\mu\Phi^A + g A_{\mu}^I
(T_I)^A_{\,\,\,B}\Phi^B
\end{equation}
Here, $\nabla_{\mu}$ denotes the ordinary space-time-covariant derivative,
$g$ is some coupling constant, and the $(T_I)^A_{\,\,\,B}$ are the
representation matrices for the matter fields $\Phi^A$. If the gauge group
is non-Abelian, there are, in addition, self-couplings among the vector
fields $A_{\mu}^I$.  A supergravity theory without such `gauge' couplings
is generally termed `ungauged'~\footnote{The terms `gauged' and `ungauged'
supergravity are only used for theories in which the supergravity sector
and the gauge sector show a certain degree of entanglement. This typically
happens when the supergravity multiplet contains vector fields that are
candidates for gauge fields. A prominent example for which this is
\emph{not} the case is four-dimensional $\mathcal{N}=1$ supergravity
coupled to $\mathcal{N}=1$ super-Yang-Mills theory with or without chiral
matter multiplets, as in the minimally supersymmetric extension of the
Standard Model. In four and five dimensions, one needs at least eight
supercharges for the supergravity multiplet to contain at least one vector
field, so that the term `gauged supergravity' is commonly used in these
dimensions only for theories with $\mathcal{N}\geq 2$ supersymmetry.}.

Typically, the local gauge symmetry of a gauged supergravity theory
reduces to a global, i.e., rigid, symmetry of an underlying ungauged
supergravity theory when the gauge coupling $g$ is turned off. In these
cases, one can iteratively construct the gauged supergravity theories from
their ungauged relatives via the Noether procedure. To this end, one first
selects a `gaugeable' subgroup, $K$, of the total global symmetry group,
$G$, of the underlying ungauged Lagrangian. One then covariantizes the
relevant derivatives {\it \`{a} la} (\ref{covder}), so as to turn the
former \emph{global} symmetry group $K$ into a \emph{local} gauge
symmetry. This typically breaks supersymmetry, but, if the gauge group $K$
was appropriately chosen, supersymmetry can be restored by adding a few
additional terms to the Lagrangian and the transformation laws. 

In this Section we recall the appropriate criteria for a group $K\subset G
$ to be gaugeable in the context of five-dimensional $\mathcal{N}=2$
supergravity theories. In the remainder of this paper we then look
for solutions to these constraints.

\subsection{General Formalism}

The minimal amount of supersymmetry in five space-time dimensions
corresponds to eight real supercharges, and is generally referred to as
$\mathcal{N}=2$ supersymmetry. The $R$-symmetry group of the underlying
Poincar\'{e} superalgebra is $USp(2)_R\cong SU(2)_R$.  The
five-dimensional $\mathcal{N}=2$ supergravity multiplet can be coupled to
vector multiplets, self-dual tensor multiplets and hypermultiplets. The
field contents of these multiplets are as follows~\footnote{Our space-time
conventions coincide with those of~\cite{GST1,GST2,GZ1,CD}, i.e., all
fermions are symplectic Majorana spinors, the metric signature is
$(-,+,+,+,+)$, and $\mu,\nu\ldots$ and $m,n,\ldots$ denote curved and flat
space-time indices, respectively.}.

\begin{itemize}
\item The supergravity multiplet
\begin{displaymath}
(e_{\mu}^m,\psi_{\mu}^i,A_{\mu})
\end{displaymath}
contains the f\"{u}nfbein $e_{\mu}^m$, an $SU(2)_R$ doublet of
gravitini $\psi_{\mu}^i$: $i=1,2$ and a vector field $A_{\mu}$.

\item A vector multiplet
\begin{displaymath}
(A_{\mu},\lambda^i,\varphi)
\end{displaymath}
consists of a vector field $A_{\mu}$, an $SU(2)_R$ doublet of spin-1/2
gaugino fermions $\lambda^i$: $i =1,2$ and one real
scalar field $\varphi$.

\item A tensor  multiplet
has the same field content as a vector multiplet, but with the vector
field $A_{\mu}$ replaced by a two-form field $B_{\mu\nu}$ satisfying
odd-dimensional duality as explained below.

\item A hypermultiplet
\begin{displaymath}
(\zeta^A,q^X)
\end{displaymath}
comprises two spin-1/2 fermions (hyperini) $\zeta^A$:
$A=1,2$, and four real scalar fields $q^X$: $X=1,\ldots,4$.
The hyperini are inert under $SU(2)_R$, which is why we have
not used the $SU(2)_R$ doublet index $i$ for these fields.
\end{itemize}

When the theory is \emph{ungauged}, vector and tensor fields can always be
dualized into each other and are physically equivalent, so one does not
have to distinguish between vector and tensor multiplets at the level of
the ungauged theory. However, this equivalence between vector and tensor
multiplets does not hold for certain \emph{gauged} theories, as we discuss
in more detail below. 

The ungauged coupling of $n$ vector and $m$ hypermultiplets to
supergravity was worked out in~\cite{GST1,Sier}. The bosonic
sector of such a theory consists of
\begin{itemize}
\item the f\"{u}nfbein $e_{\mu}^m$,
\item $(n+1)$ vector fields $A_{\mu}^{\tilde{I}}$:
$\tilde{I},\tilde{J}\ldots=0,1,\ldots,n$, where we have combined
the graviphoton with the $n$ vector fields from the $n$ vector
multiplets to form a single $(n+1)$-plet of vector fields,
\item $n$ scalar fields $\varphi^{x}$: $x,y,\ldots=1,\ldots, n$
from the $n$ vector multiplets,
\item $4m$ scalar fields $q^X$: $X,Y,\ldots=1,\ldots,4m$ from the
$m$ hypermultiplets,
\end{itemize}
The $(n+4m)$ scalar fields $\{\varphi^x,q^X\}$ parametrize a
Riemannian manifold $\mathcal{M}$ of (real) dimension $(n+4m)$,
which was found to factorize~\cite{Sier}:
\begin{equation}
\mathcal{M}=\mathcal{M}_{VS}\times \mathcal{M}_Q,
\end{equation}
where $\mathcal{M}_{VS}$ is an $n$-dimensional real manifold~\cite{GST1},
which is `very special' in a sense defined below and parametrized by the
scalar fields $\varphi^x$, and $\mathcal{M}_Q$ denotes a quaternionic
manifold of real dimension $4m$ parametrized by the hyperscalars $q^X$
\cite{bagwit}.

Introducing the Maxwell-type field strengths
$F_{\mu\nu}^{\tilde{I}}\equiv 2
\partial_{[\mu}A_{\nu]}^{\tilde{I}}$, the bosonic part of the Lagrangian 
reads~\cite{GST1,Sier}
\begin{eqnarray}\label{Lagrange}
e^{-1}\mathcal{L}_{bosonic}&=&-\frac{1}{2}R- \frac{1}{4}
{\stackrel{\circ}{a}}_{\tilde{I}\tilde{J}}
F_{\mu\nu}^{\tilde{I}}F^{\mu\nu\tilde{J}}-     \frac{1}{2}
g_{\tilde{x}\tilde{y}}(\partial_{\mu}\varphi^{\tilde{x}})(\partial^{\mu}
\varphi^{\tilde{y}})\nonumber\\
&&-\frac{1}{2} h_{XY}(\partial_{\mu}q^{X})(\partial^{\mu} q^{Y})
+\frac{e^{-1}}{6\sqrt{6}}C_{\tilde{I}\tilde{J}\tilde{K}}
\varepsilon^{\mu\nu\rho\sigma\lambda}F_{\mu\nu}^{\tilde{I}}
F_{\rho\sigma}^{\tilde{J}}A_{\lambda}^{\tilde{K}}.
\end{eqnarray}
Here, $e\equiv \det(e_{\mu}^{m})$, whereas $g_{xy} (\varphi)$ and $h_{XY}(q)$ denote, 
respectively, the  metrics on the scalar manifolds $\mathcal{M}_{VS}$ and 
$\mathcal{M}_Q$. The quantity ${\stackrel{\circ}{a}}_{\tilde{I}\tilde{J}}(\varphi)$ is 
symmetric in its indices and depends on the scalar fields $\varphi^{x}$. The completely 
symmetric tensor $C_{\tilde{I}\tilde{J}\tilde{K}}$, by contrast, is \emph{constant}, 
i.e., it does \emph{not} depend on any of the scalar fields. Because of this, the 
Lagrangian is invariant under the Maxwell-type transformations
\begin{equation}\label{Maxwell2}
A_{\mu}^{\tilde{I}}\longrightarrow A_{\mu}^{\tilde{I}} +
\partial_{\mu} \Lambda^{\tilde{I}}
\end{equation}
even though $A^{\tilde{I}}_{\mu}$ appears explicitly in the $F\wedge F
\wedge A $ term in (\ref{Lagrange}). Despite this invariance, the
above theories are still referred to as `ungauged', as we
discussed at the beginning of this Section.

The tensor $C_{\ti\tj\tk}$ turns out to determine completely the
part of the Lagrangian that is due to the supergravity and the
vector multiplets~\cite{GST1}. In particular, it completely
determines the metric of the `very special' manifold $\M_{VS}$.
To be more explicit, the $C_{\ti\tj\tk}$ define a cubic polynomial
\begin{equation}
N(h):=C_{\ti\tj\tk}h^{\ti}h^{\tj}h^{\tk}
\end{equation}
in $(n+1)$ real variables $h^{\ti}$: $\ti=0,\ldots, n$, which
endows ${\mathbb{R}}^{(n+1)}$ with the  metric
\begin{equation}\label{aij}
a_{\tilde{I}\tilde{J}}(h):=-\frac{1}{3}\frac{\partial}{\partial
h^{\tilde{I}}} \frac{\partial}{\partial h^{\tilde{J}}} \ln N(h).
\end{equation}
The $n$-dimensional `very special' manifold $\M_{VS}$ can then be
represented as the hypersurface~\cite{GST1}
\begin{equation}
N(h)=C_{\ti\tj\tk}h^{\ti}h^{\tj}h^{\tk}=1
\end{equation}
with the metric $g_{xy}$ on $\M_{VS}$ being the induced metric of this hypersurface in 
the ``ambient'' space with the metric (\ref{aij}), and furthermore we have
${\stackrel{\circ}{a}}_{\ti\tj}(\varphi)=a_{\ti\tj}|_{N=1}$.

\subsection{The Global Symmetries and their Possible Gaugings}

In this subsection we give a general overview of the different
types of global symmetries of the ungauged Lagrangian
(\ref{Lagrange}), and give a pre-classification of the possible
types of gaugings.

\subsubsection{Case I: No Hypermultiplets}

We first consider theories without hypermultiplets, which we
also describe as `Maxwell-Einstein supergravity
theories' (MESGTs). In these cases, the $C_{\ti\tj\tk}$ determine
the entire theory, and any (infinitesimal) linear transformation
\begin{eqnarray}
h^{\ti}&\longrightarrow& M^{\ti}_{\tj}h^{\tj}\label{htrafo}\\
A_{\mu}^{\ti}&\longrightarrow&
M^{\ti}_{\tj}A_{\mu}^{\tj}\label{Atrafo}
\end{eqnarray}
that leaves the $C_{\ti\tj\tk}$ invariant:
\begin{equation}\label{Cinv}
M^{\ti'}_{\,\,(\ti} C_{\tj\tk)\ti'}=0,
\end{equation}
extends to a \emph{global} symmetry of the entire Lagrangian. We call
$G_{VS}$ the
group generated   by  all  these symmetry transformations, i.e., the
invariance group of the cubic polynomial $N(h)$.
The group $G_{VS}$ has to be a subgroup of the isometry group,
$Iso(\M_{VS})$, of the scalar manifold $\M_{VS}$, which becomes
manifest if one rewrites the kinetic term of the scalar fields as
\cite{dWvP1,GST1}
\begin{displaymath}
-\frac{1}{2}g_{xy}(\partial_{\mu}\varphi^{x})(\partial^{\mu}
\varphi^{y})= \frac{3}{2}C_{\ti\tj\tk}h^{\ti}\partial_{\mu}h^{\tj}
\partial^{\mu}h^{\tk} |_{N = 1}.
\end{displaymath}
In most cases, $G_{VS}$ and $Iso(\M_{VS})$ are the same, but there
are some counterexamples~\cite{dWvP1,GST3} in which some isometries
of $\M_{VS}$ do not extend to global symmetries of the full
Lagrangian, i.e., to  symmetries of the $C_{\ti\tj\tk}$. In such
cases, it is then necessary to distinguish between  the invariance
group of the pure scalar sector, $Iso(\M_{VS})$,  and the symmetry
group of the entire Lagrangian, $G_{VS}$, because only the latter
can be gauged.

Regardless of the possible existence of this geometric symmetry
group $G_{VS}$ (for generic $C_{\ti\tj\tk}$,  $G_{VS}$ might very
well be trivial), every MESGT is in any case invariant under
global transformations of the $R$-symmetry group $SU(2)_{R}$. As
mentioned at the beginning of this Section, $SU(2)_{R}$
acts only on the indices $i$ of the fermions, not on the
`geometric' indices $ (\ti, x)$. As a consequence, the total
\emph{global} symmetry group of a MESGT factorizes:
\begin{displaymath}
\textrm{Global invariance group of a MESGT}= G_{VS}\times
SU(2)_{R}.
\end{displaymath}
On quite general grounds, one thus
obtains the following list of conceivable types of gauge
groups~\cite{GST2,GZ1,GZ3}:
\begin{itemize}
\item $U(1)_{R}\subset SU(2)_{R}$,
\item $K\subset G_{VS}$,
\item $U(1)_{R}\times K$,
\item $SU(2)_{R}\times K$ with $K\supset SU(2)$.
\end{itemize}

Here, $K$ denotes some `gaugeable' subgroup of $G_{VS}$ (see below). The
gauging of $U(1)_{R}$ turns out to be a necessary prerequisite for
obtaining Anti-de Sitter ground states~\cite{GST2,GZ1,GZ2}. On the other
hand, the gauging of $U(1)_R$ does not interfere with the gauging of a
subgroup $K$ of $G_{VS}$~\cite{GZ1}~\footnote{ We should point out one
subtle point in this regard. The gauge field of $U(1)_R$ must be a linear
combination of those vector fields that are singlets of $K$.}. This is no
longer true if one wants to gauge the \emph{entire} $R$-symmetry group
$SU(2)_{R}$, which \emph{requires} the simultaneous gauging of a subgroup
$K\subset G_{VS}$ that itself contains an $SU(2)$ subgroup $SU(2)\subset
K$~\cite{GZ3}. From this it follows that the non-trivial part of a more
explicit gauge group classification lies in the classification of the
possible gauge groups $K\subset G_{VS}$. 

What are the constraints on such gauge groups $K$? According to
(\ref{Atrafo}), the $(n+1)$ vector fields $A_{\mu}^{\ti}$
transform in a (not necessarily irreducible) $(n+1)$-dimensional
representation of the global invariance group $G_{VS}$. The
minimal consistency requirement for a subgroup $K\subset G_{VS}$
to be gaugeable is therefore that this $(n+1)$-dimensional
representation contains the adjoint of $K$ as a subrepresentation.
In the most general case, one therefore has the decomposition
\footnote{For $K$ Abelian the adjoint of $K$ and the $K$-singlets
should be identified.}:
\begin{equation}
(n+1)_{G_{VS}}\longrightarrow \textrm{adj}(K) \oplus
\textrm{singlets}(K) \oplus \textrm{non-singlets}(K).
\end{equation}

Two cases have to be distinguished:\\

(i) When the above decomposition contains no non-singlets of $K$
beyond the adjoint, it was shown in~\cite{GST2} that the gauging
can always be performed and that the resulting theory has no
scalar potential, unless one also gauges $U(1)_R$~\cite{GZ1} or
$SU(2)_R$~\cite{GZ3} in addition to $K$.\\

(ii) If, on the other hand, non-singlets beyond the adjoint do occur,
the corresponding non-singlet vector fields have to be converted
to self-dual tensor fields $B_{\mu\nu}$ in order
for the gauging to be compatible with supersymmetry~\cite{GZ1}. At
the linearized level, these tensor fields fulfill a first-order
field equation of the form~\cite{PTvN}
\begin{equation}
dB=im \ast B,
\end{equation}
where $\ast$ denotes the Hodge dual, $m$ is a massive parameter
proportional to the gauge coupling $g$, and all internal indices
have been suppressed for simplicity. Because of this equation, the
two-form fields $B_{\mu\nu}$ are no longer equivalent to vector
fields when the gauge coupling is non-zero.

For later reference, we split the index $\ti$ according to
\begin{equation}
\ti=(I,M),
\end{equation}
where $I,J,K, \ldots=1,\ldots n_{V}$ collectively denote the
vector fields in the adjoint as well as the $K$-singlets, and the
$M,N,P,\ldots =1,\ldots, n_{T}$ refer to the non-singlets of $K$,
i.e., the tensor fields.

The presence of self-dual tensor fields introduces two important
new features into the theory:
\begin{itemize}

\item Consistency with supersymmetry now requires the existence
of a non-vanishing
scalar potential, $P^{(T)}$, which can be written in the form~\cite{GZ1}
\begin{equation}\label{pot}
P^{(T)}=\frac{3}{4}g_{xy}K^{x}_I K^{y}_J h^I h^J,
\end{equation}
where the $K_J^x$ denote the Killing vectors on $\M_{VS}$ corresponding to the subgroup 
$K\subset G_{VS}\subset Iso(\M_{VS})$ of its isometry group~\footnote{As mentioned 
earlier, and contrary to what happens in four dimensions~\cite{dWvP0,ABCDFF}, this 
potential vanishes when no tensor fields are present. This can be seen directly from 
(\ref{pot}), taking into account the fact that the very special geomety of $\M_{VS}$ 
implies~\cite{GST2} that $ K^{x}_{\ti}  h^{\ti}=0$ when the summation goes over the 
\emph{full} set of indices $\ti$.}. This potential is manifestly positive definite and 
hence  can not lead to AdS ground states, unless one also gauges $U(1)_R$~\cite{GZ2}.

\item The presence of the tensor fields
implies  several new restrictions on the $C_{\ti\tj\tk}$ and the
admissible gauge groups $K\subset G_{VS}$ \cite{GZ1}.
Supersymmetry now demands that the
coefficients of the type $C_{MNP}$ and $ C_{IJM}$ have to vanish:
\begin{equation}\label{vanishingC}
C_{MNP}= C_{IJM}=0.
\end{equation}

Furthermore, the transformation matrices $\Lambda^M_{IN}$ of the
non-singlets have to be
\begin{equation}\label{LambdaC}
\Lambda^N_{IM}=\frac{2}{\sqrt{6}}\Omega^{NP}C_{MPI}\Longleftrightarrow
\Omega_{NP}\Lambda^{P}_{IM}=\frac{2}{\sqrt{6}}C_{MNI},
\end{equation}
where $\Omega_{MN}$ and $\Omega^{MN}$ are antisymmetric and
inverse to each other:
\begin{displaymath}
\Omega_{PN} \Omega^{NM}=\delta_{P}^{M}.
\end{displaymath}
For the inverse $\Omega^{MN}$ to exist, $n_{T}$ obviously has to
be even. The symmetry of the $C_{IMN}$ and equation (\ref{LambdaC})
further imply
\begin{equation}\label{Lambdasymplectic}
\Lambda^{P}_{IN}\Omega_{PM} +\Omega_{NP}\Lambda^{P}_{IM}=0 \quad
\textrm{or} \quad \Lambda_I^T\cdot\Omega + \Omega\cdot
\Lambda_I=0,
\end{equation}
i.e., the non-singlets have to transform in a \emph{symplectic}
representation of the gauge group $K$ \cite{GZ1}.
\end{itemize}

In Section 3, we exploit these restrictions and
classify those $C_{\ti\tj\tk}$ that meet all these requirements.
Having physical applications in mind, however, we only
consider \emph{compact} gauge groups $K$ that are either \\
(i) Abelian or\\
(ii) semi-simple or \\
(iii) a direct product of an Abelian and a semi-simple group.

\subsubsection{Case II: The General Case with Hypermultiplets}

When hypermultiplets are present~\cite{CD,Sier}, there is an additional
 global symmetry 
group, $Iso(\M_{Q})$, the isometry group of the quaternionic target 
space $\M_{Q}$ of 
the hyperscalars \cite{bagwit}.  However, as the hypermultiplets do not 
contain any 
vector fields themselves, any gauging of the quaternionic isometries has 
to be 
`external', i.e., it has to be done  with the vector fields $A^{I}_{\mu}$ 
of the 
supergravity and/or vector multiplets.

Two cases should be distinguished (see also~\cite{CD,LOSW2,ABCDFF}).

\begin{enumerate}
\item If one wants to gauge an Abelian subgroup $K\subset Iso(\M_{Q})$,
one needs at least $\dim(K)$ vector fields, i.e.,
$n_{V}=(\dim(K)-1)$ vector multiplets. No other restriction has to
be satisfied in the vector multiplet sector.
\item
If $K\subset Iso(\M_{Q})$ is non-Abelian, one needs at 
least $n_{V}=\dim(K) $ vector 
multiplets, but now one also needs the gauge fields to 
transform in the adjoint of $K$. 
This  means that, just as in the case without hypermultiplets, $K$ now also has to be a 
gaugeable subgroup of $G_{VS}$.
\end{enumerate}

To summarize, the gauging of a given non-trivial group of quaternionic
isometries imposes the same constraints on the gaugeable subgroups of the
very special geometry as in the case without the hypermultiplets. We
therefore focus on a classification of the gaugeable isometries of the
very special geometry. Having solved that problem, the classification of
the gaugeable quaternionic isometries is then equivalent to a
classification of all isometry groups of all possible quaternionic
manifolds~\footnote{The \emph{homogeneous} quaternionic manifolds were
classified in~\cite{dWvP2}.}. A deeper understanding of this problem would
also provide information on the possible matter representations in
five-dimensional gauged supergravities, which is also important for the
reasons mentioned in the Introduction. However, this lies beyond the scope
of this paper: for some recent results, see~\cite{dWRV}.

\section{Very Special Manifolds with Gaugeable Compact Isometries}
\setcounter{equation}{0}

Our goal is to classify the cubic polynomials
\begin{displaymath}
N(h)=C_{\ti\tj\tk}h^{\ti}
h^{\tj}h^{\tk}
\end{displaymath}
that have a non-trivial invariance group, $G_{VS}$,  with a
gaugeable compact subgroup $K\subset G_{VS}$.

Our classification is constructive, in that we write down the possible
building blocks of such polynomials, i.e., of the underlying coefficients
$C_{\ti\tj\tk}$. Besides the restrictions imposed by the gauging, these
building blocks have to satisfy one additional constraint, which is
already present in the ungauged theory. This constraint has to do with the
fact that a given set of $C_{\ti\tj\tk}$ uniquely determines the tensor
${\stackrel{\circ}{a}}_{\ti\tj}$ in the kinetic term of the vector fields
as well as the metric $g_{xy}$ of the very special manifold $\M_{VS}$.
Both ${\stackrel{\circ}{a}}_{\ti\tj}$ and $g_{xy}$ have to be positive
definite in order to be physically meaningful. 

In general, it appears difficult to see when this is
the case, because of the complicated expressions one usually gets when
evaluating (\ref{aij}) on the hypersurface $N(h)=1$. 
Fortunately, however, there is a
basis of the ambient space ${\mathbb{R}}^{(n+1)}\supset \M_{VS}$,
the `canonical basis'~\cite{GST1}, in which these
positivity properties become manifest.
In this canonical basis, the $C_{\ti\tj\tk}$ take the form
\begin{eqnarray}
C_{000}&=&  1  \nn \\
C_{00i}&=& 0  \nn \\
C_{0ij}&=& -\frac{1}{2} \delta_{ij}
\label{canonical2}\\
C_{ijk}&=&\textrm{arbitrary} \nn
\end{eqnarray}
with $i,j,k,\ldots=1,\ldots,n$. 
As indicated, the coefficients of the type
$C_{ijk}$  may be chosen at will, i.e., they parametrize  the remaining 
freedom
one has in deforming the manifold $\M_{VS}$ without spoiling
the positivity properties of $g_{xy}$ and ${\stackrel{\circ}{a}}_{\ti\tj}$.

In the above basis, the invariance condition (\ref{Cinv})
\begin{equation}
M^{\ti'}_{\,\,\, (\ti }C_{\tj\tk)\ti'}=0
\end{equation}
restricts the transformation matrices $M^{\ti'}_{\,\,\, \ti }$ to
be of the form (see also~\cite{dWvP2}):
\begin{eqnarray}
M^{0}_{\,\,\, 0 }&=&0 \nn \\
M^{i}_{\,\,\, 0 }&=&M^{0}_{\,\,\, i }\\
M^{i}_{\,\,\, j }&=&S_{ij}+A_{ij}, \nn 
\end{eqnarray}
where $S_{ij}$ is symmetric in $i$ and $j$, and $A_{ij}$ is
antisymmetric. The matrix $S_{ij}$ is given by
\begin{equation}
S_{ij}=M^{k}_{\,\,\, 0 }C_{kij},
\end{equation}
whereas $A_{ij}$ is subject to the constraint
\begin{equation}\label{AM}
C_{l(ij}A_{k)l}=M^{m}_{\,\,\, 0}\left[
C_{lm(i}C_{jk)l}-\frac{1}{2}\delta_{m(i}\delta_{jk)}\right].
\end{equation}
We are only interested in \emph{compact} symmetries of the
$C_{\ti\tj\tk}$. These are generated by the antisymmetric part of
$M^{\ti}_{\,\,\, \tj }$, i.e., we have to set $M^{i}_{\,\,\, 0
}=M^{0}_{\,\,\, i }=0$ and are left with
\begin{equation}
M^{\ti}_{\,\,\, \tj }=\left(
\begin{array}{cc}
0&0\\
0&A_{ij}
\end{array}
\right)
\end{equation}
with
\begin{eqnarray}
A_{ij}&=&- A_{ji} \Longleftrightarrow A_{ij}\in \mathfrak{so}(n)\\
C_{l(ij}A_{k)l}&=&0.
\end{eqnarray}
Hence, a compact symmetry group of the cubic polynomial $N(h)$ is
given by the subgroup of the $SO(n)$ rotations of the $h^{i}$ that
also leave the coefficients $C_{ijk}$ invariant~\footnote{This
also implies that the action of a compact gauge group $K\subset
G_{VS}\subset Iso(\M_{VS})$  has always at least one fixed point
on $\M_{VS}$, namely the `base point'~\cite{GST1}
$h_{c}^{\ti}=(1, 0, \ldots, 0)\in \M\subset
\mathbb{R}^{n+1}$, which is left invariant under the action of
$SO(n)\supset K$. This in turn guarantees the existence of at
least one critical point of the potential
$P^{(T)}$ related to the tensor fields, because $K_{I}^{x}=0$ at this point 
- see (\ref{pot}). Obviously, this
critical point corresponds to a Minkowski ground state of the
theory (unless $U(1)_R$ is also gauged~\cite{GZ2}), and it can be shown
that this ground state is
$\mathcal{N}=2$ supersymmetric.}. All we have to do then is to
classify the possible $C_{ijk}$ that preserve \emph{gaugeable}
subgroups $K$ of this $SO(n)$.

\subsection{The Most Symmetric Case: $C_{ijk}=0$}

We start this classification with the simplest case
\begin{equation}\label{Cijk=0}
C_{ijk}=0
\end{equation}
for all $i,j,k,\ldots=1,\ldots,n$. In this most symmetric case, the
polynomial $N(h)$ is obviously invariant under the full $SO(n)$. In fact,
it is easy to see that (\ref{Cijk=0}) automatically implies
$M^{0}_{\,\,\,i}=M^{i}_{\,\,\,0}=0$ via the constraint (\ref{AM}), i.e.,
there are no non-compact symmetries, and $SO(n)$ is the full symmetry
group of $N(h)$. It is interesting to note that the manifolds based on
(\ref{Cijk=0}) are in general not homogeneous, i.e., they are not
contained in the classification of homogeneous very special manifolds
given in~\cite{dWvP2}. Their peculiar geometry can best be seen by
introducing the following `radial coordinate' for the scalar manifold 
\[ r^2 = \frac{3}{2} \sum_{i=1}^{n} h^ih^i. \] 
The hypersurface condition then takes
the form 
\[ N= h^0 [(h^0)^2 -r^2]=1, \] 
which can be rewritten in terms of the
`lightcone' coordinates $ r_{\pm}=\frac{1}{2} ( h^0\pm r) $ as 
\[ r_+ r_- (r_+ + r_- )=4. \] 
This hypersurface has two disconnected components . The
topology of each connected component of the full hypersurface is of the
form 
\[ \M_{VS}= \aleph \times S^{n-1}, \] 
where $\aleph$ is the surface in the $(h^0,r)$ plane given by $N=1$.

We now turn to the gaugeable subgroups of $G_{VS}=SO(n)$. The
components $h^{i}$ transform in the $\mathbf{n}$ of $SO(n)$. Any gaugeable
compact subgroup $K\subset G_{VS}$ must therefore be a subgroup of $SO(n)$
such that the adjoint representation of $K$ is contained in the
$\mathbf{n}$ of $SO(n)$. However, the adjoint of \emph{any} compact group
$K$ is always embeddable in the defining representation of any $SO(n)$
with $n\geq \dim{(K)}$, because the positive-definite Cartan-Killing form
$\kappa_{ab}$ provides an invariant metric for the adjoint of $K$. Hence,
\emph{any} compact group $K$ with $\dim{(K)}\leq n$ can be gauged if
(\ref{Cijk=0})  holds. If $n-\dim{(K)}=:r>0$, one has $(r+1)$ spectator
vector fields, one of them being $A_{\mu}^{0}$, which can be identified
with the graviphoton. By construction, the other $\dim{(K)}$ vector fields
transform in the adjoint of $K$ and act as $K$-gauge fields. The spectator
vector fields can in principle be used to gauge also $U(1)_R$ and/or
Abelian isometries of the hyperscalar manifold $\M_{Q}$, if they exist.

Note that the gaugings described above do not introduce any tensor fields.
The only way to obtain a theory with tensor fields in the above model is
by gauging an $SO(2)$ subgroup of $SO(n)$: $n \geq 2$, with $A_{\mu}^{0}$
being the $SO(2)$ gauge field.  This follows because the transformation
matrices $\Lambda_{IN}^M$ of such tensor fields would have to be related
to some $C_{IMN}$ via (\ref{LambdaC}). In the case at hand, i.e., with
$C_{ijk}=0$, such coefficients could only come from the $C_{0ij}$ with
$I=0$ - see (\ref{canonical2}). Thus $A_{\mu}^0$ would be the only vector
field that could couple to such tensor fields, and the latter can only be
charged with respect to a single $SO(2)$ subgroup of $SO(n)$. 

We discuss such Abelian gaugings with tensor fields in a slightly more
general context in Section 3.3.

We now consider cubic polynomials $N(h)$ with \emph{non}-trivial
$C_{ijk}$. These polynomials can be viewed as deformations of the simplest
case (\ref{Cijk=0}). Since there are no completely symmetric invariant
tensors of rank three in the $\mathbf{n}$ of $SO(n)$, such deformations
will in general break $SO(n)$ to a subgroup. We are only interested in the
case where this surviving symmetry group (or a subgroup thereof) can be
gauged. As usual, we refer to this gaugeable subgroup of $SO(n)$ as $K$.
Note also that, whereas the case $C_{ijk}=0$ does not in general lead to
homogeneous spaces, some of the deformations with $C_{ijk}\neq 0$ do. 

\subsection{Nontrivial $C_{ijk}$ without Tensor Fields}

We first consider the case where the gauging of $K$ does not
involve tensor fields. In this case, the $\mathbf{n}$ of $SO(n)$
decomposes according to
\begin{displaymath}
\mathbf{n}=\textrm{adjoint}(K)\oplus \textrm{singlets}(K).
\end{displaymath}
Assuming the above decomposition, an Abelian factor of $K$ could
not act non-trivially on anything. Thus, when no tensor fields are
present, a compact gauge group
$K\subset G_{VS}$ has to be semi-simple~\footnote{Of course, one
could still gauge
$U(1)_R$ and/or an Abelian  subgroup of $Iso(\M_Q)$ in addition to
$K\subset G_{VS}$.}.

We split the indices $i=1,\ldots, n$ as follows:
\begin{equation}
i=(a,\alpha),
\end{equation}
where $a,b,\ldots=1,\ldots, p\equiv \dim{(K)}$ correspond to the
adjoint of $K$, and $\alpha,\beta,\ldots=1,\ldots,r$ label the $r$
singlets, where $p+r=n$.

Before we proceed, we note that the term of the form
\begin{displaymath}
C_{0ij}h^{0}h^{i}h^{j}=-\frac{1}{2} h^{0}\delta_{ij}h^{i}h^{j}
\end{displaymath}
appearing in the canonical basis (\ref{canonical2}) now
reads
\begin{equation}\label{deltaij}
C_{0ij}h^{0}h^{i}h^{j}=-\frac{1}{2}h^{0}(\delta_{ab}h^{a}h^{b}
+\delta_{\alpha\beta}h^{\alpha}h^{\beta}).
\end{equation}

Our goal is to find all possible deformations of the relation
$C_{ijk}=0$ (\ref{Cijk=0}) that are consistent with the
invariance under $K$. Clearly, coefficients of the form
$C_{a\alpha\beta}$ transform in the adjoint of $K$ and can
therefore never be invariant under $K$ transformations when $K$ is
semi-simple. Indeed, any such non-trivial $C_{a\alpha\beta}$ would
correspond to an Abelian ideal of $K$, in contradiction to the assumption
of semi-simplicity. Hence, we have
\begin{equation}
C_{a\alpha\beta}=0.
\end{equation}
It remains to discuss the coefficients of the following forms.

\begin{enumerate}
\item $C_{\alpha\beta\gamma}$:\\
Since the $h^{\alpha}$ are $K$-singlets, any
$C_{\alpha\beta\gamma}$ are consistent with $K$ invariance.

\item $C_{\alpha a b}$:\\
In order to be invariant under $K$, $C_{\alpha a b}$ has to be an
invariant symmetric tensor of rank 2 of the adjoint representation
of $K$. The only such object is the Cartan-Killing form
$\kappa_{ab}$ of $K$. However, in order for the $\delta_{ab}$ term
in (\ref{deltaij}) to be invariant under $K$, one has to work in a
basis where $\kappa_{ab}=\delta_{ab}$, so that any term $C_{\alpha
a b}$ must be of the form
\begin{displaymath}
C_{\alpha ab}=c_{\alpha}\delta_{ab}
\end{displaymath}
with some arbitrary constants $c_{\alpha}$.

\item $C_{abc}$:\\
In order for this term to be invariant under the action of $K$, it
has to be equal to a completely symmetric invariant tensor of rank
3 of the adjoint representation of $K$. As was already emphasized
in~\cite{GZ1}, such tensors exist only for the groups $SU(N)$ with
$N\geq 3$ (or  products thereof), where they are given by the
Gell-Mann $d$ symbols:
\begin{displaymath}
d_{abc}=\textrm{Tr}(T_{a}\{T_{b},T_{c}\})
\end{displaymath}
with the $T_{a}$ being the generators of $SU(N)$. Hence, if
$K=SU(N)$: $N\geq 3$, or if $K$ is a product of such $SU(N)$
factors, a term $C_{abc}=d_{abc}$ can be
introduced without spoiling the $K$ invariance of the cubic
polynomial $N(h)$. As an interesting side remark, we note that an
$SU(N)$ gauging with $C_{abc}=d_{abc}$ leads to a quantization
condition for the gauge coupling constant of $K$~\cite{GST1.5},
whereas an $SU(N)$ gauging with $C_{abc}=0$ does not. The reason
for this difference is the non-triviality of the Chern-Simons term
in the case $C_{abc}=d_{abc}$: see \cite{GST1.5} for further
details.

\end{enumerate}

\subsection{Non-Trivial $C_{ijk}$ with Tensor Fields}

Before we start with the classification of the possible $C_{ijk}$,
we first prove the following
\vspace{2mm} \\
\textbf{Observation:} If tensor fields are present, a compact
gauge group
$K$ has to have at least one Abelian factor.\\
\emph{Proof:} We first recall that a compact group $K\subset G_{VS}$
can act non-trivially only on the $h^{i}$: $i=1,\ldots, n$, i.e.,
$h^{0}$ has to be inert under $K$. Hence, all the tensor field
indices
$M,N,\ldots=1,\ldots, 2m\equiv n_{T}$ must be among the
$i,j,\ldots=1,\ldots, n$. We therefore split the index $i$ as
follows
\begin{displaymath}
i=(I',M),
\end{displaymath}
where the indices $I',J',\ldots=1,\ldots, (n-2m)$ label the vector
fields $A_{\mu}^{i}$ that are \emph{not} dualized to tensor
fields. The total set of vector fields that survive the tensor
field dualization $A_{\mu}^{M}\rightarrow B_{\mu\nu}^{M}$ is thus
given by
\begin{displaymath}
A_{\mu}^{I}=(A_{\mu}^{0},A_{\mu}^{I'}).
\end{displaymath}
We recall that the $h^{M}$ transform as follows (cf. 
(\ref{htrafo})) under $K$:
\begin{displaymath}
h^{M}\longmapsto \Lambda_{IN}^{M}h^{N},
\end{displaymath}
with
\begin{equation}\label{CLambda2}
\Lambda_{IM}^{N}=\frac{2}{\sqrt{6}}\Omega^{NP}C_{IMP}.
\end{equation}
being the $K$ transformation matrices of the tensor fields
$B_{\mu\nu}^{M}$.
Furthermore, we note that the term
\begin{displaymath}
C_{0ij}h^{0}h^{i}h^{j}=-\frac{1}{2} h^{0}\delta_{ij}h^{i}h^{j}
\end{displaymath}
appearing in the canonical basis (\ref{canonical2}) contains the
term
\begin{equation}\label{deltaij5}
C_{0MN}h^{0}h^{M}h^{N}=-\frac{1}{2}h^{0}\delta_{MN}h^{M}h^{N}.
\end{equation}
The presence of this term has two important consequences:
\begin{enumerate}
\item There is always a non-vanishing matrix $\Lambda_{0M}^{N}$ given
by (\ref{CLambda2}), which, in the case at hand, becomes
\begin{equation}\label{LisOmega}
\Lambda_{0}=-\frac{1}{\sqrt{6}}\Omega^{-1}.
\end{equation}

\item Since $h^{0}$ is inert under $K$, the $K$ invariance of the term
(\ref{deltaij5}) requires the matrices $\Lambda_{IM}^{N}$ to be
orthogonal:
\begin{equation}\label{Lisorthogonal}
\Lambda_{I}^{T}+\Lambda_{I}=0.
\end{equation}
\end{enumerate}

Recalling that the
 $\Lambda_{IM}^{N}$ also have to be symplectic (\ref{Lambdasymplectic}):
\begin{equation}\label{Lissymplectic}
\Lambda_{I}^{T}\cdot\Omega +\Omega \cdot \Lambda_{I}=0,
\end{equation}
we have
\begin{eqnarray*}
\Omega\cdot
[\Lambda_{0},\Lambda_{I}]\cdot\Omega&\stackrel{(\ref{LisOmega})}{=}&-
\frac{1}{
\sqrt{6}}
[\Lambda_{I}\cdot \Omega-\Omega\cdot\Lambda_{I}]\\
&\stackrel{(\ref{Lisorthogonal})}{=}& 
\frac{1}{\sqrt{6}}[\Lambda_{I}^{T}\cdot\Omega +\Omega \cdot \Lambda_{I}]\\
&\stackrel{(\ref{Lissymplectic})}{=}&0,
\end{eqnarray*}
i.e., the (non-trivial) matrix $\Lambda_{0}$ commutes with all the
$\Lambda_{I}$, and $K$ has to have at least one Abelian factor,
which acts nontrivially on the tensor fields via
$\Lambda_{0N}^{M}$.
\hfill \emph{Q.E.D.}\\
\vspace{1mm}\\

As a corollary of (\ref{Lisorthogonal}) and
(\ref{Lissymplectic}), we note that, choosing
$\Omega=i\sigma_{2}\otimes \mathbf{1}_{m}$, each matrix
$\Lambda_{I}$ has to be of the form
\begin{equation}\label{Lambdamatrix}
\Lambda=\left(
\begin{array}{cc}
X&Y\\
-Y&X
\end{array}\right)\textrm{ with } \left\{
\begin{array}{c}
X=-X^{T}\\
Y=Y^{T},
\end{array}\right.
\end{equation}
where $X$ and $Y$ are real $(m\times m)$-matrices. Obviously,
$X+iY$ is anti-Hermitian, i.e., an element of $\mathfrak{u}(m)$
(the above is nothing but the standard embedding of
$\mathfrak{u}(m)$ into $\mathfrak{sp}(2m,\mathbb{R})$ or
$\mathfrak{so}(2m)$). This already shows that
the allowed representation matrices
$\Lambda_{IM}^{N}$, and hence the allowed coefficients
$C_{IMN}$, are in one-to-one correspondence with \emph{unitary}
$m$-dimensional representations of the compact gauge group $K$.

We now return to our classification of the possible
coefficients $C_{ijk}$ in the presence of tensor fields. Due to
the above Observation, $K$ has to have at least one Abelian
factor. We first cover the case when $K=K'\times U(1)$ with
$K'$ semi-simple, and then the case when $K=U(1)^{l}$ is purely
Abelian. The most general case is then obtained by rather obvious
combinations.

\subsubsection{$K=K'\times U(1)$}

We first assume $K=K'\times U(1)$ with $K'$ semi-simple and
with both factors acting non-trivially on the same set of tensor
fields. The $\textbf{n}$ of $SO(n)$ then decomposes with respect
to $K'$ as
\begin{displaymath}
\mathbf{n}=\textrm{adjoint}(K')\oplus \textrm{singlets}(K')\oplus
\textrm{non-singlets}(K'),
\end{displaymath}
where, by assumption, the $U(1)$ factor acts non-trivially only on
the non-singlets of $K'$. Consequently, we split the index
$i=1,\ldots, n$ into three subsets of indices:
\begin{equation}
i=(a,\alpha,M),
\end{equation}
where, $a,b,\ldots=1,\ldots, p\equiv \dim{(K')}$ correspond to the
adjoint of $K'$; $\alpha,\beta,\ldots=1,\ldots,r$ label the $r$
singlets; and $M,N,\ldots=1,\ldots, 2m$ refer to the $2m$
non-singlets: $p+r+2m=n$.

As explained in Section 2.2.1, the presence of the non-singlets
$h^{M}$ requires the conversion of the corresponding vector fields
$A_{\mu}^{M}$ to antisymmetric tensor fields $B_{\mu\nu}^{M}$. For
consistency, the coefficients of the form $C_{IJM}$ and $C_{MNP}$
then have to vanish (see (\ref{vanishingC})).
Recalling that, in our current notation, the index $I$ comprises
the indices $(0,a,\alpha)$, the set of possibly non-vanishing
coefficients $C_{ijk}$ therefore shrinks to
\begin{displaymath}
C_{\alpha\beta\gamma}, \quad C_{\alpha ab}, \quad C_{\alpha \beta
a}, \quad C_{abc}, \quad C_{a MN}, \quad C_{\alpha MN}.
\end{displaymath}
The allowed $C_{ijk}$ are constrained by the requirement that they
be invariant under $K$. The coefficients of the type $C_{\alpha\beta a}$
are $U(1)$ singlets, but they transform in the adjoint of $K'$ and
can therefore never contain any singlets of $K'$ when $K'$ is
semi-simple (see above). Hence,
\begin{displaymath}
C_{\alpha\beta a}=0,
\end{displaymath}
and we are left with the following.
\begin{enumerate}
\item $C_{\alpha \beta \gamma}$:\\
Any coefficient of this type would automatically be inert under $K$, and
can therefore have any arbitrary value.
\item $C_{\alpha ab}$:\\
This term is a $U(1)$ singlet. As explained in our discussion of
the corresponding term for the case without tensor fields, the
only possible form of this term consistent with invariance under
$K'$ is
\begin{displaymath}
C_{\alpha ab}=c_{\alpha}\delta_{ab},
\end{displaymath}
with arbitrary constants $c_{\alpha}$.

\item $C_{abc}$\\
As explained earlier, this term can be either zero or equal to the
$d$ symbols of $SU(N)$, if $K'=SU(N)$: $N\geq 3$, or if $K'$ is a
product of such $SU(N)$ factors.

\item $C_{aMN}$:\\
We first note that, in general,
any term of the form $C_{IMN}$ with $I\in \{ 0,a,\alpha\}$
is automatically invariant under $K$. In fact, under a $K$
transformation, it transforms as
\begin{displaymath}
C_{IMN}\longmapsto
f_{JI}^{K}C_{KMN}+\Lambda_{JM}^{P}C_{IPN}+\Lambda_{JN}^{P}C_{IMP},
\end{displaymath}
which vanishes automatically because of relation (\ref{CLambda2})
and the fact that the $\Lambda_{IM}^{N}$ generate a representation
of $K$:
\begin{equation}\label{commutator}
[\Lambda_{I},\Lambda_{J}]=\Lambda_{K}f_{IJ}^{K}.
\end{equation}

The  $C_{aMN}$ are uniquely determined by the $\Lambda_{aM}^{N}$
via (\ref{CLambda2}). All we have to do then is  to classify the
possible $K'$ representation matrices $\Lambda_{aM}^{N}$. From our
discussion around (\ref{Lambdamatrix}), however, it follows
that the possible $\Lambda_{aM}^{N}$ are in one-to-one
correspondence with $m$-dimensional unitary representations of
$K'$. Since $K'$ is compact, \emph{any} representation
of $K'$ can be chosen to be unitary, and any such unitary
representation can be embedded into $(2m\times 2m)$ matrices of
the form (\ref{Lambdamatrix}) to form a possible set of
$\Lambda_{aM}^{N}$ or, equivalently, a possible set of $C_{aMN}$.

\item  $C_{\alpha MN}$:\\
The $C_{\alpha MN}$ also give rise to transformation matrices
$\Lambda_{\alpha M}^{N}$ via (\ref{CLambda2}). Since, by
assumption, our gauge group is $K=K'\times U(1)$, and the $\Lambda_{aM}^{N}$
already generate $K'$, the $\Lambda_{\alpha M}^{N}$ are either
zero or they correspond to the $U(1)$ factor. However, we already know
that the (non-vanishing) matrix
$\Lambda_{0M}^{N}$ generates this $U(1)$ factor - see the proof at
the beginning of this subsection. Since we assumed only one
$U(1)$ factor, the $\Lambda_{\alpha M}^{N}$ can be at most
proportional to $\Lambda_{0M}^{N}$, otherwise they would give rise
to another, independent, Abelian factor in the gauge group $K$.
For the $C_{\alpha MN}$ this means that they can be at most
(remember that $C_{0MN}=-(1/2)\delta_{MN}$)
\begin{displaymath}
    C_{\alpha MN}=d_{\alpha}\delta_{MN}
\end{displaymath}
for some constants $d_{\alpha}$. In that case, the $U(1)$ gauge
field would be the linear combination
\begin{displaymath}
A_{\mu}[U(1)]=\left[-\frac{1}{2}A_{\mu}^{0}+d_{\alpha}
        A_{\mu}^{\alpha}\right].
\end{displaymath}

\end{enumerate}

\subsubsection{$K=U(1)^l$}

We now come to the case when $K=U(1)^{l}$ is purely Abelian.  We assume
for simplicity that all the $U(1)$ factors act on the same set of tensor
fields. If there were Abelian groups acting on mutually disjoint sets of
tensor fields, the cubic polynomial would simply decompose into several
subpieces of the type to be described below. 

Assuming now the above gauge group structure,
the $\mathbf{n}$ of $SO(n)$ decomposes as follows:
\begin{displaymath}
    \mathbf{n}=\textrm{singlets}(K)\oplus
    \textrm{non-singlets}(K).
\end{displaymath}
We denote the singlets of $K$ by $\alpha,\beta,\ldots=1,\ldots,r$ and
the non-singlets by $M,N,\ldots =1,\ldots, 2m$, i.e., we split
\begin{displaymath}
    i=(\alpha,M).
\end{displaymath}
The possible non-vanishing $C_{ijk}$ are now the following.
\begin{enumerate}
\item $C_{\alpha\beta\gamma}$:\\
These coefficients are automatically singlets of $K$, and can
therefore be chosen arbitrarily.

\item $C_{\alpha MN}$:\\
Via (\ref{CLambda2}), these
coefficients are related to the $K$-transformation
matrices $\Lambda_{\alpha M}^{N}$, which are again of the
form (\ref{Lambdamatrix}). The maximal Abelian subgroup of $U(m)$
is $m$-dimensional, so that $K$ can be at most $U(1)^{m}$. In the
special case $K=U(1)$, the same arguments that were used in the
case $K=K'\times U(1)$ apply, and the $C_{\alpha MN}$ could be at
most
\begin{displaymath}
    C_{\alpha MN}=d_{\alpha}\delta_{MN}
\end{displaymath}
for some constants $d_{\alpha}$. In this case, the $U(1)$ gauge
field would again be the linear combination
\begin{displaymath}
A_{\mu}[U(1)]=\left[-\frac{1}{2}A_{\mu}^{0}+d_{\alpha}
        A_{\mu}^{\alpha}\right].
\end{displaymath}

\end{enumerate}

It is now rather straightforward to construct more general cubic
polynomials by various combinations of the above basic building
blocks.

We close this subsection with a comment on the nature of the tensor
fields. As we have seen, a compact gauge
group $K\subset G_{VS}$ has to be semi-simple when \emph{no} tensor fields
are introduced. Conversely, when tensor fields \emph{are} present, a
compact gauge group $K\subset G_{VS}$ can \emph{never} be semi-simple; it
has to contain at least one Abelian factor.  This suggests the following
interpretation.

If a compact group $K\subset G_{VS}$ is gauged, and tensor fields have to be
introduced, one has at least one $\mathcal{N}=2$ supersymmetric Minkowski
ground state of the potential $P^{(T)}$ (see the footnote on page 11). The
tensor multiplets should therefore admit an interpretation as
$\mathcal{N}=2$ Poincar\'{e} supermultiplets, at least for compact $K$. 
Since the tensor fields satisfy a massive field equation, such a multiplet
would necessarily have to be massive. This is consistent with the form of
the scalar potential $P^{(T)}$ in (\ref{pot}), which can be easily shown
to be quadratic in the $h^{M}$. Due to their $K$ transformation
properties, the $h^{M}$ have a natural interpretation as parametrizing the
scalar fields in the tensor multiplets. Thus, $P^{(T)}$ can be interpreted
as providing the mass terms for the massive scalars in the (massive) 
tensor multiplets. Such a massive tensor multiplet would have to be a
centrally-charged BPS multiplet in order to have the same number of
degrees of freedom as the massless vector multiplet from which it emerged. 
Indeed, the five-dimensional $\mathcal{N}=2$ Poincar\'{e} superalgebra
with central charges has precisely one such BPS multiplet with exactly the
right field content (see, e.g., \cite{Strathdee,PTvN}).  It is then
tempting to identify the $U(1)$ factor in the (compact)  gauge group $K$
with the (necessarily gauged) central charge of the corresponding
Poincar\'{e} superalgebra.

Note that the whole situation changes when one gauges $U(1)_R$ as well as
$K$. As shown in~\cite{GZ2}, this kind of gauging typically leads to a
$\mathcal{N} = 2$ supersymmetric AdS ground state, and the tensor
multiplets would then have a natural interpretation as the self-dual
tensor multiplets of the $\mathcal{N} = 2$ AdS superalgebra described
in~\cite{GRW}.

\section{An Illustrative Exercise: The Standard Model Gauge Group}

As an illustration of the general analysis of Section 3, we now
demonstrate how to obtain the Standard Model gauge group
$K_{SM}=SU(3)\times SU(2) \times U(1)$ within five-dimensional
supergravity.

Since the dimension of the Standard Model gauge group is
$\dim(K_{SM})=12$, we need at least twelve vector fields, i.e., at least
eleven vector multiplets in addition to the supergravity multiplet. In
addition to this minimal field content, there might be additional vector
multiplets and/or some tensor multiplets.  We first discuss the case
without any tensor multiplets.

\subsection{Case 1: No Tensor Multiplets}

When there are no tensor multiplets, all the vector fields have to
transform in the adjoint representation of $K_{SM}$, or they must be
singlets under the gauge group, as discussed in Section 2.2.1. Since the
adjoint of the $U(1)$ factor of $K_{SM}$ is trivial, this $U(1)$ factor
has to act trivially on all the vector fields. In order to obtain fields
charged under the $U(1)$ factor without introducing tensor fields, one
would therefore have to gauge a $U(1)_R$ subgroup of the $R$-symmetry
group and/or an Abelian isometry of the hypermultiplet scalar manifold
$\M_{Q}$ (provided such an isometry exists). Neither of these Abelian
gaugings interferes with the classification of the admissible very special
manifolds $\M_{VS}$. We can thus, as in Section 3.2, restrict our
attention to the semi-simple part of $K_{SM}$. 

Working in the canonical basis, the $(n+1)$ vector fields $A_{\mu}^{\ti}$
are split into 
\begin{displaymath}
A_{\mu}^{\ti}=(A_{\mu}^{0},A_{\mu}^{i})
\end{displaymath}
with $i=1,\ldots,n$ $(n\geq 11)$, and the  $C_{\ti\tj\tk}$ are of the form
\begin{eqnarray}
C_{000}&=&  1  \nn \\  C_{00i}&=& 0  \nn \\ 
 C_{0ij}&=& -\frac{1}{2} \delta_{ij}
\label{canonical3}\\
C_{ijk}&=&\textrm{not yet fixed} \nn
\end{eqnarray}
A compact symmetry group acts trivially on  $ A_{\mu}^{0}$,
so that the adjoint vector fields of $SU(2)$ and $SU(3)$ have to be recruited
from the $A_{\mu}^{i}$, which we therefore split into
\begin{displaymath}
A_{\mu}^{i}=(A_{\mu}^{\hat{a}},A_{\mu}^{\bar{a}},A_{\mu}^{\alpha}),
\end{displaymath}
where $A_{\mu}^{\hat{a}}$ and $A_{\mu}^{\bar{a}}$ denote
the adjoint vector fields of $SU(2)$ and $SU(3)$, respectively,
whereas the $A_{\mu}^{\alpha}$ stand for additional $K_{SM}$ singlets
(which may or may not be present).

As described in Section 3.2, the coefficients $C_{ijk}$ are now 
restricted by their $SU(2)\times SU(3)$ invariance to take the 
following forms:

\begin{eqnarray}
C_{\alpha\beta\gamma}&=&\textrm{arbitrary} \nn \\ 
 C_{\alpha\beta\hat{a}}&=&0 \nn \\ \nn
C_{\alpha\beta\bar{a}}&=&0\\ \nn
 C_{\alpha \hat{a}\hat{b}}&=&c_{\alpha}\delta_{\hat{a}\hat{b}}\\ \nn
  C_{\alpha \bar{a}\bar{b}}&=&d_{\alpha}\delta_{\bar{a}\bar{b}}\\ \nn 
  C_{\alpha \hat{a}\bar{b}}&=&0  \nn \\
C_{ \hat{a} \hat{b} \hat{c}}&=&0\label{SU2d}\\ 
 C_{ \hat{a} \hat{b} \bar{c}}&=&0 \nn \\ 
\nn C_{ \hat{a} \bar{b} \bar{c}}&=&0\\ \nn
 C_{ \bar{a} \bar{b} \bar{c}}&=&b d_{\bar{a} \bar{b} \bar{c}}, \nn
\end{eqnarray}
where $C_{\alpha\beta\gamma}$, 
$c_{\alpha}$, $d_{\alpha}$ and  $b$ denote some arbitrary coefficients,
and the $d_{\bar{a} \bar{b} \bar{c}}$ are the $d$ symbols of $SU(3)$.
As mentioned earlier, there is no such term for the $SU(2)$ factor - see
(\ref{SU2d}). A number of remarks are now relevant.

\noindent\textbf{Remark 1:} A linear combination of the $SU(2)\times SU(3)$ 
singlets $A_{\mu}^{0}$ and 
$A_{\mu}^{\alpha}$ could always be used to gauge $U(1)_{R}$ and/or
an Abelian isometry of the hyperscalar manifold $\M_{VS}$.
Similarly, the $SU(2)$ and the $SU(3)$ gauge fields $A_{\mu}^{\hat{a}}$
and $A_{\mu}^{\bar{a}}$ could always be used to gauge $SU(2)$ and $SU(3)$
subgroups of $Iso(\M_{Q})$, provided such subgroups exist. Depending on 
the particular quaternionic manifold one considers, one would then 
get hypermultiplets transforming in certain representations of
$K_{SM}$ (if this is what wants to have).\\
\textbf{Remark 2:} As mentioned in Section 3.3, a non-zero value 
for $b$ would lead to a 
quantization condition for the $SU(3)$ coupling constant in the sense 
described in~\cite{GST1.5}.\\
\textbf{Remark 3:} If $n$ satisfies its lower bound $n=11$, i.e.,
if there are no $A_{\mu}^{\alpha}$, and $A_{\mu}^{0}$ is the only 
$K_{SM}$ singlet, one has two options:
\begin{enumerate}
\item $b=0$:\\
corresponding to the simple case $C_{ijk}=0$ described in
Section 3.1,
\item $b\neq 0$:\\
leading to a quantization condition for the $SU(3)$ coupling
constant - see Remark 2 above.
\end{enumerate}
Thus, the minimal case $n=11$ is fairly restrictive and allows only for  
a one-parameter family of scalar manifolds $\M_{VS}$. 
The price one has to pay for this rigidity
is that the $U(1)$ factor of the Standard Model gauge group 
would have to be gauged with the only remaining vector field $A_{\mu}^{0}$,
so that \emph{all} the vector fields would have to participate in the 
gauging, including the \emph{graviphoton}. If, for some reason, 
one does not want the graviphoton to be part of the Standard Model
gauge fields, one would need at least
$n=12$, which then introduces more arbitrariness into the theory
via the new undetermined coefficients $C_{\alpha\beta\gamma}$, 
$c_{\alpha}$, $d_{\alpha}$.\\
\textbf{Remark 4:}
None of the `minimal' cases with $n=11$, described in Remark 3,
corresponds to a symmetric space $\M_{VS}$. 
In order to implement  the Standard Model 
gauge group in a model based on a \emph{symmetric} space $\M_{VS}$,
one needs $n\geq 12$, i.e., at least one additional singlet $A_{\mu}^{\alpha}$.
The corresponding values for $C_{\alpha\beta\gamma}$, 
$c_{\alpha}$, $d_{\alpha}$ and  $b$ can be read off from
equations (\ref{gJ}) and (\ref{ngJ})  in the Appendix.\\
\textbf{Remark 5:} If there are at least three $A_{\mu}^{\alpha}$,
and if the $C_{\alpha\beta\gamma}$, $c_{\alpha}$, $d_{\alpha}$ 
are chosen appropriately, i.e., 
as described in Section 3.2, one could introduce further non-Abelian 
gauge factors. Similarly -- if this is desired -- 
one could consider embedding $K_{SM}$
into larger gauge  groups like $SU(5)$, $SO(10)$ etc. 
and write out the resulting
restrictions on the $C_{\ti\tj\tk}$. We leave these extensions as
exercises.

\subsection{Case 2: The Presence of Tensor Fields}

We now consider the case with tensor fields. Self-dual tensor fields
always have to be charged under some gauge group~\cite{GZ1}. 
In our case, this group 
could simply be $K_{SM}$ itself, or some part of it. On the other hand, 
the tensor fields
could also be charged under some other gauge group factor which does not belong
to the Standard Model gauge group $K_{SM}$. In order to keep the 
degree of complexity at a minimum, we only consider the case 
when $K_{SM}$ is indeed the full gauge group, and  
the tensor fields are charged under $K_{SM}=SU(3)\times SU(2)\times U(1)$.
This is then exactly the case we considered in Section 3.3.1,
and we can simply quote the results of that Section.
As the tensor fields always come in pairs, we now need $n\geq 11+2=13$.

We again work in the canonical basis, but now 
split the index $i$ as follows
\begin{equation}
i=(\hat{a},\bar{a},\alpha,M),
\end{equation}
where $\hat{a}$ and $\bar{a}$ correspond to the adjoint of
$SU(2)$ and  $SU(3)$, respectively, whereas $\alpha$ refers to the
singlets and
$M$ to the non-singlets (i.e., the tensor fields) of $K_{SM}$.

The admissible $C_{ijk}$ are now given by (see Section 3.3.1):
\begin{eqnarray}
C_{\alpha\beta\gamma}&=&\textrm{arbitrary} \nn \\
C_{\alpha\beta\hat{a}}&=&0 \nn \\
C_{\alpha\beta\bar{a}}&=&0 \nn \\
C_{\alpha \hat{a}\hat{b}}&=&c_{\alpha}\delta_{\hat{a}\hat{b}} \nn \\
C_{\alpha \bar{a}\bar{b}}&=&d_{\alpha}\delta_{\bar{a}\bar{b}} \nn \\
C_{\alpha \hat{a}\bar{b}}&=&0 \nn \\
C_{ \hat{a} \hat{b} \hat{c}}&=&0 \nn \\
C_{ \hat{a} \hat{b} \bar{c}}&=&0 \nn \\
C_{ \hat{a} \bar{b} \bar{c}}&=&0  \\
C_{ \bar{a} \bar{b} \bar{c}}&=&b d_{\bar{a} \bar{b} \bar{c}} \nn \\
C_{M\bar{a}\bar{b}}&=&0=C_{M\hat{a}\hat{b}}=C_{M\bar{a}\hat{b}}
=C_{M\bar{a}\alpha}=C_{M\hat{a}\alpha}=C_{M\alpha\beta} \nn \\
C_{\hat{a}MN}&=&\frac{\sqrt{6}}{2}\Omega_{MP}\Lambda_{\hat{a}N}^{P} \nn \\
C_{\bar{a}MN}&=&\frac{\sqrt{6}}{2}\Omega_{MP}\Lambda_{\bar{a}N}^{P} \nn \\
C_{\alpha MN}&=& e_{\alpha}\delta_{MN} \nn \\
C_{MNP}&=&0 \nn .
\end{eqnarray}
Here, $C_{\alpha\beta\gamma}$, $c_{\alpha}$, $d_{\alpha}$, $e_{\alpha}$ and 
$b$
are again arbitrary coefficients, which might or might not be zero,
and $d_{\bar{a} \bar{b} \bar{c}}$ again stand for the $SU(3)$ $d$ symbols.
The matrices $\Lambda_{\hat{a}N}^{P}$ and $\Lambda_{\bar{a}N}^{P}$
are, respectively,
the $SU(2)$ and $SU(3)$ transformation matrices of the tensor fields.
They can be related  to $\left(\frac{n_T}{2}\right)$-dimensional 
unitary representations of $SU(2)$ 
and $SU(3)$ via (\ref{Lambdamatrix}), where $n_T$ denotes the (even)
number of tensor fields. As for the $U(1)$ factor, the tensor fields 
would transform 
via the representation matrix $\Lambda\sim \Omega^{-1}$ as in 
(\ref{LisOmega}),
with the corresponding $U(1)$ gauge field being the linear combination
\begin{displaymath}
A_{\mu}[U(1)]=\left[-\frac{1}{2}A_{\mu}^{0}+e_{\alpha}
        A_{\mu}^{\alpha}\right].
\end{displaymath}
(see the last item in Section 3.3.1).

Once again, one finds that the minimal case $n=13$ leads to a very small 
number of choices for $\M_{VS}$, and requires the graviphoton to be 
one of the Standard Model gauge fields. To be more precise, the 
coefficients $C_{\alpha\beta\gamma}$, 
$c_{\alpha}$, $d_{\alpha}$, $e_{\alpha}$ have to vanish, because there is no 
$A_{\mu}^{\alpha}$, and the $SU(2)$ and $SU(3)$ transformation matrices
$\Lambda_{\hat{a}N}^{P}$ and $\Lambda_{\bar{a}N}^{P}$ would have to vanish
because there is no non-trivial representation of these groups of the form 
(\ref{Lambdamatrix}) for the minimal case $n_T=2$: any such representation
would  be related to \emph{one}-dimensional (and hence trivial) 
unitary representations of $SU(2)$ and $SU(3)$
via (\ref{Lambdamatrix}).
Thus, in the minimal 
embedding of the Standard Model gauge group with two tensor fields,
the tensor fields form an $U(1)\cong SO(2)$ doublet and are inert under 
$SU(2)\times SU(3)$, and the only free parameter is the coefficient $b$.

Departure from the minimal
value $n=13$ then again introduces more arbitrariness into the theory 
because of the new unconstrained coefficients $C_{\alpha\beta\gamma}$, 
$c_{\alpha}$, $d_{\alpha}$, $e_{\alpha}$, which, in the absence of any 
further selection principle, can have any value.

\section{Summary and Conclusions}

We gave in the Introduction various motivations for considering the
possible gaugings of five-dimensional $\mathcal{N}=2$ supergravity. 
Whereas globally supersymmetric $\mathcal{N}=2$ Yang-Mills theories in
five dimensions can be studied for any compact gauge group without very
stringent restrictions on the field content~\cite{5Dfieldtheory}, it is
not {\it a priori} clear what new restrictions are imposed by the
non-linear structures introduced by a coupling to supergravity. Since
gravity plays an important r\^{o}le in the current interest in
five-dimensional theories, it is therefore important to analyze the
constraints local supersymmetry imposes on the gauge sector.

In general, this is a difficult geometrical problem, which helps explain
why most studies in the past focussed on theories with very peculiar
classes of scalar manifolds. In fact, almost all the known concrete
examples involved symmetric~\cite{GST1,GST2,GST3,dWvP1,GZ1} or at least
homogeneous spaces~\cite{dWvP2,GZ1}. However, thanks to the very special
geometry of the five-dimensional vector multiplet moduli space encoded in
the coefficients $C_{\ti\tj\tk}$, this geometrical problem can be reduced
to a purely algebraic one.  The entire analysis boils down to a
classification of the possible $C_{\ti\tj\tk}$ that are consistent with
invariance under the  gauge group $K$. 

We have solved this algebraic classification problem for all compact gauge
groups that are semi-simple, or Abelian, or a direct product of a
semi-simple and an Abelian group. Our algebraic approach allowed us to go
beyond the limitations set by the restriction to homogeneous or symmetric
spaces. In fact, from the viewpoint of possible gauge symmetries,
symmetric and homogeneous spaces are just particular examples of much
larger classes of possible scalar manifolds. 

Our main results can be summarized as follows.
\begin{enumerate}
\item $K$ semi-simple:\\
\emph{Any} compact semi-simple group $K$ can be gauged provided one 
respects certain constraints on the field content and on the 
couplings encoded in the $C_{\ti\tj\tk}$. These constraints can be
found in Section 3.1 and 3.2. The key results are
\begin{itemize}
\item One always needs \emph{at least} $n=\dim(K)$ vector multiplets,
i.e., there is always at least one spectator vector field which can be 
identified with the graviphoton. Note that this no longer holds true
for non-compact gauge groups. There, one can construct examples in which
\emph{all} the vector fields, including the graviphoton, act as the
gauge fields of $K$~\cite{GST2}. 
\item In the minimal case $n=\dim(K)$, the scalar manifold $\M_{VS}$ 
is \emph{fixed} whenever $K$ does not contain an $SU(N)$ factor with 
$N\geq 3$. If, on the other hand, 
$K$ does contain $SU(N)$ factors with $N \geq 3$, each such
$SU(N)$ factor gives rise to one  undetermined parameter
in the $C_{\ti\tj\tk}$ and hence in the resulting scalar manifold
$\M_{VS}$, as is illustrated by the Standard Model example discussed
in Section 4. The minimal case $n=\dim(K)$ 
does not in general lead to symmetric spaces.
\item If $K$ is purely semi-simple and compact, tensor fields are ruled out,
because they would need at least one $U(1)$ factor in the gauge group.
Again this result no longer holds true for non-compact gauge groups,
where one could also have tensors for purely semi-simple
$K$~\cite{GST2,GZ1}.
\item As a by-product of the previous item, we found a natural 
interpretation of the tensor multiplets in terms of massive BPS multiplets of the 
centrally-extended Poincar\'{e} superalgebra, and also as self-dual tensor multiplets of 
the corresponding AdS superalgebra. Which of these two interpretations applies depends 
whether one also gauges $U(1)_{R}$ or not, as we discuss at the end of Section 3.
\end{itemize}
\item $K$ Abelian:\\
There are essentially two ways to implement an Abelian gauge group $K$.
If the Abelian gauge group is $U(1)_R$ and/or an Abelian isometry
of the hypermultiplet moduli space $\M_{Q}$, no restriction on the
very special geometry of the vector multiplet sector is imposed:
the very special geometry is blind towards such gaugings.

The other possibility, which is the one we focused on in this paper,
is when the Abelian gauge group acts non-trivially on the 
very special manifold $\M_{VS}$, i.e., when one gauges an Abelian isometry 
of $\M_{VS}$. This case \emph{always} requires tensor fields charged under 
$K$.

\item $K=K_{semi-simple}\times K_{Abelian}$:\\
This is essentially a combination of (i) and (ii), so, again, if the
Abelian factor acts non-trivially on $\M_{VS}$, one must have some tensor 
fields charged under this Abelian factor. The only new feature is now that 
the tensor fields can also be charged with respect to the semi-simple part 
of the gauge group. This is an interesting difference from the analogous 
$\mathcal{N}=4$ theories~\cite{DHZ}, where the tensor fields can only 
be charged with respect to a one-dimensional Abelian group. 
As for the possible 
$K$ representations of the tensor fields, we found that they are in 
one-to-one corresponence with unitary representations of $K$.
\end{enumerate}

In this paper, we have provided five-dimensional model-builders with a
necessary toolkit, enabling them to construct the most general theory with
any given gauge group. As an example of such a construction, we considered
the Standard Model gauge group as a toy model in Section 4. 

The matter content allowable in a general five-dimensional $\mathcal{N}=2$
supergravity theory requires a further discussion of the hypermultiplet
sector, which goes beyond the scope of this paper. Another worthwhile
extension of the present work would be to consider the analogous
classification problem for gaugings of six-dimensional supergravity. There
is increasing interest in six-dimensional models of particle physics: 
see~\cite{HNOS} and references therein. So far, phenomenological
constructions have not incorporated explicitly the constraints that would
be imposed by local supersymmetry in six dimensions~\cite{sixd}, which are
even stronger than those in five dimensions. 

We foresee a fruitful continuation of the dialogue between model-building
and explorations of the structures of higher-dimensional supergravity
theories.

{\bf Acknowledgements:} M.G. acknowledges the hospitality of the 
Caltech-USC Institute 
and the IAS at Princeton, and  M.Z. the hospitality of CERN Theory 
Division, while part 
of this work was done. The work of M.Z. was supported by the
``Schwerpunktprogramm 1094'' of the DFG.

\vspace{0.5cm}

\noindent
{\bf\Large Appendix}

\begin{appendix}
\section{Gauge Theories in Families of Symmetric Spaces}

\renewcommand{\theequation}{A.\arabic{equation}}

%\appendix

%\section{Gauge Theories in Families of Symmetric Spaces}
%\setcounter{equation}{0}

As an illustration of the more abstract discussion in Section 3,
we show in this Appendix how to recover some well-known examples in the
language used in that Section. These examples correspond to the
scalar manifolds
\begin{itemize}
\item $\M_{VS}=SO(1,1)\times SO(n-1,1)/SO(n-1)$:\\
 (the `generic Jordan family'~\cite{GST1})
\item $\M_{VS}=SO(n,1)/SO(n)$\\
 (the `generic non-Jordan family'~\cite{GST3})
\item $\M_{VS}=SL(3,\mathbb{C})/SU(3)$,
\end{itemize}
which, apart from three additional cousins of the last one,
exhaust all the very special manifolds that are symmetric
spaces~\cite{GST3,dWvP1}.

\subsection{ $\M_{VS}=SO(1,1)\times SO(n-1,1)/SO(n-1)$}
In the canonical basis, the corresponding cubic polynomial is
given by
\begin{equation}
N(h)=\left[(h^{0})^{3}-\frac{3}{2}h^{0}
\delta_{ij}h^{i}h^{j}-\frac{1}{\sqrt{2}}(h^{1})^{3}
+\frac{3}{\sqrt{2}} h^{1}[(h^{2})^{2}+\ldots +(h^{n})^{2}]\right].
\label{gJ}
\end{equation}
In terms of the framework in Section 3, this polynomial can be
interpreted in different ways, depending on which group $K$ one
chooses as the gauge group. Using indices
\begin{eqnarray}
\alpha&=& 1\nonumber\\
a&=&2,\ldots,n,\nonumber
\end{eqnarray}
for example, it could correspond to one of the theories where a
semi-simple group $K \subset SO(n-1)\subset SO(n)$ with
adjoint$(K)\subset \mathbf{(n-1)}\subset \mathbf{n}$ can be gauged
without the introduction of tensor fields, as in Section 3.2.

However, one can also interpret the indices $\{2,\ldots n\}$ (or a
subset thereof) as tensor field indices $M,N\ldots$. This would
then correspond to an $SO(2) \subset SO(n)$ gauging with tensor
fields, with the $SO(2)$ gauge field being proportional to the
linear combination $[A_{\mu}^{0}-\sqrt{2}A_{\mu}^{1}]$, as in
Section 3.3.2.

Other interpretations involving combinations of the above are of
course also possible. This illustrates    that, in general,  for one and
the same manifold $\M_{VS}$, various different types of gaugings are possible,
and, conversely,  that the $C_{\ti\tj\tk}$ we constructed in Sections 3.2
and 3.3 might sometimes describe the same manifold $\M_{VS}$.

We note finally that
the transformation $h^{\ti}\mapsto {\tilde{h}}^{\ti}$ with
\begin{eqnarray}
{\tilde{h}}^{0}&=&\frac{1}{\sqrt{3}}[h^{0}-\sqrt{2}h^{1}]\nonumber\\
{\tilde{h}}^{1}&=&\frac{1}{\sqrt{3}}[\sqrt{2}h^{0}+h^{1}]\nonumber\\
{\tilde{h}}^{\ti}&=&h^{\ti} \textrm{ for } \ti=2,\ldots n\nonumber
\end{eqnarray}
leads to the following simple form 
\begin{displaymath}
N(\tilde{h})=\left(\frac{3}{2}\right)^{\frac{3}{2}}\left( 
\sqrt{2}{\tilde{h}}^{0}
[({\tilde{h}}^{1})^{2}-
({\tilde{h}}^{2})^{2}-\ldots -({\tilde{h}}^{n})^{2}]\right),
\end{displaymath}
which is no longer in the canonical
basis, but makes the full non-compact symmetry
$Iso(\M_{VS})=G_{VS}=SO(1,1)\times SO(n-1,1)$ manifest.

\subsection{$\M_{VS}=SO(n,1)/SO(n)$}

In the canonical basis, the corresponding cubic polynomial reads
\begin{equation}
N(h)=\left[(h^{0})^{3}-\frac{3}{2}h^{0}
\delta_{ij}h^{i}h^{j}+\frac{1}{\sqrt{2}}(h^{1})^{3}
+\frac{3}{2\sqrt{2}} h^{1}[(h^{2})^{2}+\ldots
+(h^{n})^{2}]\right].
\label{ngJ}
\end{equation}
This is, apart from two (important) prefactors, of the same form
as the polynomials of the generic Jordan family. Therefore, the
discussion of the possible compact  gauge groups $K$ is very
similar and is not repeated here. Giving up the canonical
basis, the above polynomial can also be simplified by a coordinate
transformation similar to that described for the generic Jordan
family. The definition
\begin{eqnarray}
{\tilde{h}}^{0}&=&\frac{1}{\sqrt{3}}[h^{0}+\sqrt{2}h^{1}]\nonumber\\
{\tilde{h}}^{1}&=&\frac{1}{\sqrt{3}}[\sqrt{2}h^{0}-h^{1}]\nonumber\\
{\tilde{h}}^{\ti}&=&h^{\ti} \textrm{ for } \ti=2,\ldots n\nonumber
\end{eqnarray}
leads to
\begin{displaymath}
N(\tilde{h})=\left(\frac{3}{2}\right)^{\frac{3}{2}}\left(
\sqrt{2}{\tilde{h}}^{0}
({\tilde{h}}^{1})^{2}-
{\tilde{h}}^{1}[({\tilde{h}}^{2})^{2}+\ldots
+({\tilde{h}}^{n})^{2}]\right).
\end{displaymath}
We note that not all isometries of the scalar manifolds $\M_{VS}$ in this
family are symmetries of the full $N=2$ supergravity \cite{dWvP1}. As
stressed earlier, only the subgroup of the isometry group that leaves
$N(\tilde{h})$ invariant gets extended to a symmetry group of the full
supergravity. In this case it turns out to be the $(n-1)$-dimensional
Euclidean subgroup of $SO(n,1)$. 

\subsection{$\M=SL(3,\mathbb{C})/SU(3)$}

In this model, which corresponds to the Jordan algebra,
$J_{3}^{\mathbb{C}}$, of complex Hermitian $(3\times3)$ matrices
\cite{GST1}, the index $i$ runs from $1$ to $8$.
We first  decompose this index according to
$i = (a,4,M)$ with
\begin{eqnarray}
a,b,\ldots&=&1,\ldots,3\nonumber\\
M,N,\ldots&=&5,\ldots,8\nonumber.
\end{eqnarray}
In the canonical basis, the underlying cubic polynomial can then
be written as
\begin{eqnarray}
N(h)&=&\left[(h^{0})^{3}-\frac{3}{2}
h^{0}\delta_{ij}h^{i}h^{j}+\frac{3}{\sqrt{2}}h^{4}[\delta_{ab}h^{a}h^{b}-
\frac{1}{2}\delta_{MN}
h^{M}h^{N}]\right.\nonumber\\
& &\left.-\frac{1}{\sqrt{2}}(h^{4})^{3}+\left(\frac{3}{2}\right)^{3/2}
\gamma_{aMN} h^{a}h^{M}h^{N}\right],\label{complexmagical}
\end{eqnarray}
where
\begin{eqnarray}
\gamma_{1}&=&\mathbf{1}_{2}\otimes \sigma_{1}\nonumber\\
\gamma_{2}&=&-\sigma_{2}\otimes\sigma_{2}\nonumber\\
\gamma_{3}&=&\mathbf{1}_{2}\otimes \sigma_{3}\nonumber.
\end{eqnarray}
This form makes it easy to verify  that one can gauge an
$SU(2)\times U(1)$ group acting nontrivially on a set of four
tensor fields $B_{\mu\nu}^{M}$, as in Section 3.3.1.

The $SU(2)$ vector fields are $A_{\mu}^{a}$, and the $U(1)$ gauge
field is proportional to the linear combination $[\sqrt{2}
A_{\mu}^{0}+A_{\mu}^{4}]$. This kind of gauging was examined in
detail in~\cite{GZ3}.

On the other hand, the above polynomial can also be understood
differently. After some relabelling, one finds that the above
polynomial is just
\begin{displaymath}
N(h)=\left[(h^{0})^{3}-\frac{3}{2}h^{0}
\delta_{ij}h^{i}h^{j}+d_{ijk}h^{i}h^{j}h^{k}\right],
\end{displaymath}
where  $i,j,\ldots=1,\ldots, 8$, with the $d_{ijk}$ being the
$d$ symbols of $SU(3)$. In this form, it becomes obvious that one
can also gauge $SU(3)$ without introducing any tensor fields, as
in~\cite{GST2} and our discussion in Section 3.2.

Finally, a  somewhat more concise form of (\ref{complexmagical})
is obtained via  a transformation to the new coordinates 
\begin{eqnarray}
{\tilde{h}}^{0}&=&\frac{1}{\sqrt{3}}(\sqrt{2}h^{0}+h^{4})\nonumber\\
{\tilde{h}}^{4}&=&\frac{1}{\sqrt{3}}(h^{0}-\sqrt{2}h^{4})\nonumber\\
{\tilde{h}}^{\ti}&=&h^{\ti} \textrm{ for } \ti\neq 0,4,\nonumber
\end{eqnarray}
which no longer
correspond to the canonical basis. In terms of these,
\begin{equation}\label{complexeta}
N(\tilde{h})=\left(\frac{3}{2}\right)^{\frac{3}{2}}\left(\sqrt{2}{\tilde{h}}^{4}
\eta_{\alpha\beta}{\tilde{h}}^{
\alpha}{\tilde{h}}^{\beta}+\gamma_{\alpha
MN}{\tilde{h}}^{\alpha}{\tilde{h}}^{M}{\tilde{h}}^{N}\right),
\end{equation}
where
\begin{eqnarray}
\alpha, \beta,\ldots &=& 0,1,2,3\nonumber\\
\eta_{\alpha \beta}&=&\textrm{diag}(+,-,-,-)\nonumber\\
\gamma_{0}&=&-\mathbf{1}_{4}\nonumber.
\end{eqnarray}
This is the parametrization used in~\cite{GST1}. Indeed, it is now
easy to verify that (\ref{complexeta}) is nothing but the
determinant of
\begin{displaymath}
\tilde{\mathbf{h}}=\left(\frac{3}{2}\right)^{\frac{1}{2}}\left(
\begin{array}{ccc}
\sqrt{2}{\tilde{h}}^{4}&{\tilde{h}}^{5}-i{\tilde{h}}^{7}&
{\tilde{h}}^{6}-i
{\tilde{h}}^{8}\\
{\tilde{h}}^{5}+i{\tilde{h}}^{7}&{\tilde{h}}^{0}+{\tilde{h}}^{3}&
{\tilde{h}}^{1}-i{\tilde{h}}^{2}\\
{\tilde{h}}^{6}+i{\tilde{h}}^{8}&{\tilde{h}}^{1}+i{\tilde{h}}^{2}&
{\tilde{h}}^{0}-{\tilde{h}}^{3}
\end{array}\right),
\end{displaymath}
i.e., the determinant of an  element $\tilde{\mathbf{h}}$ of the Jordan
algebra $J_{3}^{\mathbb{C}}$ of complex Hermitian $(3\times
3)$-matrices \cite{GST1}.

\end{appendix}

\end{document}